\documentclass{INTERSPEECH2023}
\usepackage{dialogue}
\usepackage{caption}
\usepackage{subcaption}

\interspeechcameraready

\title{Automatic Evaluation of Turn-taking Cues in \\Conversational Speech Synthesis}
\name{Erik Ekstedt, Siyang Wang, Éva Székely, Joakim Gustafson, Gabriel Skantze}
\address{KTH, Royal Institute of Technology, Stockholm, Sweden}
\email{erikekst@kth.se, siyangw@kth.se, szekely@kth.se, jkgu@kth.se, skantze@kth.se}

\begin{document}
\maketitle
 
\begin{abstract}
Turn-taking is a fundamental aspect of human communication where speakers convey their intention to either hold, or yield, their turn through prosodic cues. Using the recently proposed Voice Activity Projection model, we propose an automatic evaluation approach to measure these aspects for conversational speech synthesis. We investigate the ability of three commercial, and two open-source, Text-To-Speech (TTS) systems ability to generate turn-taking cues over simulated turns. By varying the stimuli, or controlling the prosody, we analyze the models performances. We show that while commercial TTS largely provide appropriate cues, they often produce ambiguous signals, and that further improvements are possible. TTS, trained on read or spontaneous speech, produce strong turn-hold but weak turn-yield cues. We argue that this approach, that focus on functional aspects of interaction, provides a useful addition to other important speech metrics, such as intelligibility and naturalness.
\end{abstract}

\section{Introduction}
In recent years, there has a been an increased interest in developing conversational Text-to-Speech (TTS) \cite{o2022combining, guo2021conversational,  raitio2020controllable, szekely2019spontaneous}. Whereas earlier conversational systems had to rely on TTS built for read speech, conversational TTS will allow  systems to interact in a more natural and fluid way, more closely resembling a human-human conversation. Many commercial vendors are now also offering conversational variants of their voices. 

Until recently, the evaluation of conversational TTS has been primarily centered around the voice's perceived naturalness and intelligibility, but this approach is being progressively challenged \cite{wagner2019speech}. 
While these aspects are important, we think there are other, more functional, aspects that also should be considered, especially in a conversational setting. 
One such aspect is the way in which the voice may help to coordinate the interaction between the participants, more specifically turn-taking. 
It is important to model turn-taking accurately in spoken dialog systems (SDS), in order to avoid long response delays or inadvertent interruptions (from both the user and the system). The failure of appropriate timing can substantially deteriorate the quality of the interaction \cite{Skantze2021}. 
It is well known that humans send and receive various turn-yielding and turn-holding cues to coordinate turn-taking \cite{sacks:74, duncan100829, Skantze2021}. For example, a syntactically or semantically incomplete phrase may signal a turn-hold, whereas a complete phrase may be turn-yielding \cite{ford_thompson:96}. A filled pause is a strong cue to turn-hold \cite{clark100117, Ball1975}. When syntax and semantics is ambiguous, prosody, gestures, or gaze can also be informative \cite{duncan100829}. With regards to prosody, flat pitch, higher intensity and longer duration are associated with turn-hold, whereas a rising or falling pitch, lower intensity and shorter duration are associated with yielding the turn. These cues allow humans to take turns with very small gaps (around 200ms), while avoiding large overlaps \cite{levinson_torreira:15}. 

The modeling of turn-taking in conversational systems has so far mostly focused on detecting turn-taking cues in the user's voice, in order to allow the system to take turns or give backchannels at appropriate places \cite{Skantze2021}. However, it is equally important that the system's voice also exhibits accurate cues, so that the user knows when to take the turn and when to allow the system to finish its turn. This is problematic in current TTS systems, as there is typically no control over these cues. Thus, there is a risk that the user might interrupt the system at the wrong places (unintended barge-in), or that the user in other ways gets confused over the allocation of the floor. 

To evaluate turn-taking cues in TTS, one option could be to ask human raters to listen to TTS samples and 
ask them to press a button when they expect a turn-shift, similar to psycholinguistic experiments on understanding turn-taking cues in human speech \cite{ruiter:06,bogels_torreira:15}. However, this is clearly a costly method and not feasible for large-scale evaluations. Thus, an automatic method would be desirable, that could complement other automatic evaluation metrics, such as MOSNet \cite{williams2020comparison} and ASR \cite{taylor2021confidence}.

In this paper, we introduce an automatic method for evaluating turn-taking cues in conversational TTS, based on Voice Activity Projection (VAP), a turn-taking model proposed by \cite{vap}. 
The model has been shown to outperform prior work \cite{skantze2017} that demonstrated that computational turn-taking models trained in a self-supervised fashion perform better than humans on predicting the next speaker on recorded data. Prior work has also shown that the VAP model is sensitive to prosodic cues in synthesized speech \cite{vapProsody}, as well as filled pauses \cite{jiang2023}. 
Here, we use the model to assess the likelihood that a user would take the turn at each frame in a synthesized spoken utterance. This way, we can make sure that this likelihood is as low as possible while the system is supposed to have the turn (i.e., it has more things to say before yielding), while it should be as high as possible towards the end of the turn. While we only focus on offline evaluation here, such a model could potentially also be used as a training objective when developing the TTS. The VAP model\footnote{\url{https://github.com/ErikEkstedt/VoiceActivityProjection}} and listening samples\footnote{\url{https://erikekstedt.github.io/vap_tts}} are publically available. 

\section{Automatic Evaluation Method}
\begin{figure}[t]
  \centering
  \includegraphics[width=.8\linewidth]{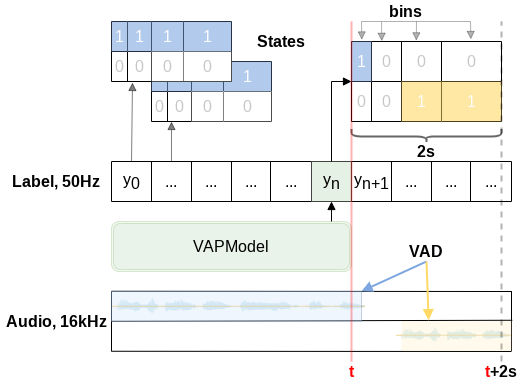}
  \caption{The VAP model receives a stereo channel audio input and predicts a discrete label, y, at frame, n, corresponding to time, t. Each label represents a state that consists of 8 binary bins, 4 for each speaker, spanning the next 2s of dialog. 
  }
  \label{fig:vap}
\end{figure}

Voice Activity Projection \cite{vap} (VAP) is a training objective where the voice activity (VA) of two speakers are predicted incrementally (left-to-right) over the course of a dialog. The VA is defined in binary terms (speech/no-speech), and the two speakers’ future activities are jointly encoded into a discrete state that represents the upcoming 2s of dialog.
The states are defined by discretizing the 2s windows of activity into eight smaller sub-state-bins, four for each speaker, of increasing duration (0.2s, 0.4s, 0.6s, 0.8s) and are considered active if they contain a majority of VA. This discretization step result in $2^8 = 256$ possible discrete states (labels) to predict during training, see Figure \ref{fig:vap}.

The VAP model used in this experiment is a stereo version of the original VAP model that operates on two separate waveforms (one for each speaker) and is trained on the Fisher part 1 and the Switchboard corpus~\cite{swb, fisher}. The model consists of a CPC-encoder \cite{cpc, cpc_across} that extracts framewise representations from the raw audio followed by a 4-layer transformer \cite{transformer} decoder with cross-attention between the two speaker channels (5.79M parameters).

During inference we scale each sub-state-bin, with their associated label probability, and combine all contributions to a single aggregate state representation.
We define two separate probabilities representing the next speaker predictions inside the most immediate region, $P_{now}$ (0-\SI{600}{\milli\second}), and the more distant future, $P_{fut}$ (600-\SI{2000}{\milli\second}), i.e. the first and last four sub-state-bins, respectively. 
The probabilities are normalized across speakers to produce a final value between 0 and 1 that represent the prediction probability of speaker A being active in the corresponding region.

\begin{figure}[t]
    \centering
    \includegraphics[width=\linewidth]{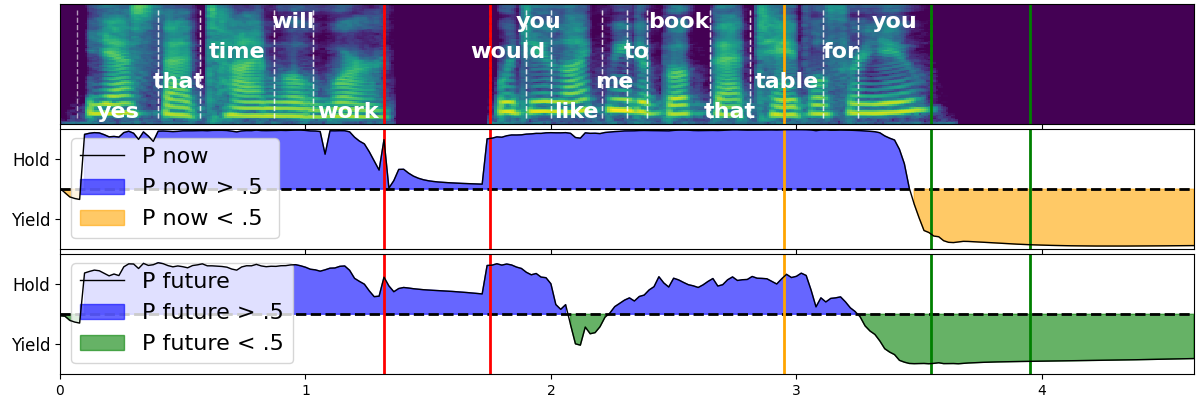}
    \caption{Example (Amazon TTS):  "Yes that time will work. Would you like me to book it for you?". From the top, we show the mel-spectrogram, the $P_{now}$, and $P_{fut}$ values. The dashed black line signifies equal speaker probability, with blue areas (hold) above and yellow or green (yield) below. Vertical lines mark the pause (red), Early Yield (orange, green), and Late Yield (green) zones. The model labels the pause as a Weak and Strong Hold and the end as an Early and Late Yield.}
    \label{fig:example}
\end{figure}

For our automatic evaluation method, we use the VAP-model to assess these predictions during pauses, when the system should hold the turn, and towards the end, when the system should yield the turn. In this way, the VAP model works as as a user model. 
In the more specific scenario we investigate in this paper, we focus on turns containing two sentences -- a statement and a question -- which will naturally include a pause that can be mistaken for a turn-shift and can cause the user to ``interrupt'' the system, if the TTS generates ambiguous turn-yielding signals. 
Furthermore, we are interested in whether the user would not just be able to detect that the turn has been yielded, but also \textit{predict} that the turn is about to be yielded (see Figure \ref{fig:example} for an example of this). 

Given the $P_{now/fut}$ values over the generated turns we define four metrics covering different regions. 
First, we focus on the pause between the sentences (red vertical lines) and consider it a \textbf{Weak Hold} if the long term prediction $P_{fut}$ favors the agent. It's considered weak because it allows for the user to be the most likely speaker in the short term ($P_{now}$), corresponding to an invitation of a quick back-and-forth, like a user backchannel or acknowledgment \cite{Yngve:70}. Second, we define a \textbf{Strong Hold} to be the subset where both the $P_{now}$ and $P_{fut}$ values favors the agent. Third, we focus on the last \SI{600}{\milli\second} of speech (between the orange and the first green line) and define it as an \textbf{Early Yield} if $P_{fut}$ favors the user. Lastly, the silence after the turn (green lines) is a \textbf{Late Yield} if both $P_{now}$ and $P_{fut}$ predicts user activity.
Arguably, the desired outcome for a conversational system planning to say two sentences would be to send clear turn-holding signals before the pause, while signaling that the turn is about to be yielded towards the end. 

\section{Experiment}
We consider a hypothetical spoken dialog system (SDS) setup where a TTS model is used to generate speech in a task-oriented setting. To simulate this setup we extract text data from the MultiWoZ corpus \cite{multiwoz}. The corpus contains over 10,000 annotated written dialogs covering 8 different domains where two humans have a fictional interaction where one has the role of a ``user'' and the other a ''clerk``, here referred to as the agent. The users have goals and several sub-goals to complete over the course of a dialog, such as booking a hotel, a taxi, a train or a restaurant, among others.

We extract agent turns that consist of a sentence pair where a statement, \textbf{SD}, (ending with a period) is followed by a question, \textbf{Q}, (ending with a question mark). 
For simplicity, and to avoid unexpected behavior, we omit sentence pairs that includes commas, digits, contains less than 5 words per sentence, and where the total amount of characters are less than 50 or more than 250. 
Finally, we only keep sentence pairs where both sentences end with a word containing a single syllable, in order to simplify the prosody manipulation in section \ref{sec:results}. 
In total, we extract 1482 sentence pairs, such as the one shown in Figure~\ref{fig:example}.

\subsection{Text-To-Speech}
We utilize three popular ASR services, namely Amazon\footnote{\url{https://aws.amazon.com/polly/}}, Google\footnote{\url{https://cloud.google.com/text-to-speech}} and Microsoft\footnote{\url{https://azure.microsoft.com/en-us/products/cognitive-services/text-to-speech}} and use their most natural and conversational american english neural TTS engines: ``Joanna-neural-en-US'' (Amazon), ``en-US-Neural2-C'' (Google) and ``en-US-JennyNeural'' (Microsoft). Additionally, we include the open-source FastPitch (\textbf{FP}) model\footnote{\url{https://github.com/NVIDIA/DeepLearningExamples/tree/master/PyTorch/SpeechSynthesis/FastPitch}} \cite{lancucki2021fastpitch} (46.27M parameters), trained on the LJSpeech\footnote{\url{https://keithito.com/LJ-Speech-Dataset/}} corpus, using the pre-trained checkpoints available to the speech community at large. We also trained a conversational TTS on the widely known Tacotron 2 (\textbf{TT2}) architecture \cite{shen2018natural} (28.21M parameters), with modifications from \cite{valle2020mellotron} that allow for control of duration and pitch at the phoneme and word level. 
The voice is trained on the ThinkComputers Corpus \cite{szekely2019spontaneous}, a corpus created from the recordings of a podcast which is made available in the public domain\footnote{\url{https://archive.org/details/podcasts_miscellaneous} Creator: ThinkComputers}.  
For the FastPitch and modified Tacotron 2 systems, the speech signal is decoded using the neural vocoder HiFi-GAN \cite{kong2020hifi} (13.93M parameters).

\subsection{Results}
\label{sec:results}

We generate speech for the extracted turns and normalize the pause (and the Late Yield) duration to be \SI{400}{\milli\second}, using forced alignment \cite{forced_aligner}, to focus on the prosodic signals of the speech rather than varying lengths of the pauses. 
After the silence normalization we apply the VAP model and extract the proposed metrics over the regions of interest. Additionally, we approximate the intelligibility and conceived naturalness of the generated speech using the open-source ASR model Whisper\footnote{\url{https://github.com/openai/whisper}} (large) to extract the word error rate, WER, and the automatic MOS predictor \cite{cooper2022generalization} that output a score between 1 and 5. 
 
All systems signal a Weak Hold over the vast majority of pauses, where TT2 achieves the highest score of 97\% and Google the lowest score of 79\%, see Table \ref{tab:metrics_orig}. A lower Weak Hold score raises the risk that a system prematurely signals the end of its turn. 
For the Strong Hold metric, TT2 outperforms the other systems, producing 93\% Strong Holds, while Amazon and FP produce around 30\%, and Google and Microsoft achieve around 20\%. All systems except TT2 have a high risk of inviting user activity inside the pause, which, if not handled correctly by the SDS, could result in a dialog breakdown. 
As discussed earlier, a Weak Hold might be acceptable (or even desirable, if the user is invited to give a backchannel) for an interactive conversational system. This means that context of the SDS interaction have to be taken into consideration \cite{wagner2019speech} and factored in to the analysis of the importance of the Strong Hold metric in general.

\begin{table}[th]
  \caption{The aggregate metrics for the ORIGINAL speech. All values are percentages (\%) except for the MOS.}
  \label{tab:metrics_orig}
  \centering
  \begin{tabular}{lccccc}
    \toprule
    \textbf{Metric}&\textbf{AMZN}&\textbf{GGL}&\textbf{MSFT}&\textbf{FP}&\textbf{TT2}\\ 
    \midrule
    Weak Hold$\uparrow$& 87 & 79 & 86 & 84 & \textbf{97} \\
    Strong Hold$\uparrow$& 31 & 21 & 20 & 30 & \textbf{93} \\
    Early Yield$\uparrow$& \textbf{44} & 28 & 32 & 5 & 4 \\
    Late Yield $\uparrow$& \textbf{95} & \textbf{95} & 90 & 39 & 14 \\
    \midrule
    MOS$\uparrow$& 4.2 & 4.3 & \textbf{4.8} & 4.4 & 3.9 \\
    WER $\downarrow$& 2.7 & 2.7 & \textbf{2.3} & 5.6 & 5.2\\
    \bottomrule
  \end{tabular}
\end{table}
Towards the end of the turns, Amazon outperforms all other systems and provides Early Yield signals for 44\% of the samples. During an interaction this could allow for better conversational flow by preparing the users for their expected upcoming turn. After the end of the generated speech, all commercial systems signal a Late Yield for 90-95\% of the samples. This means that while the systems may not always provide early prosodic turn-yield signals, they do convey their intention to yield the turn upon its completion.
However, both TT2 and FP heavily under-perform on this metric and provide 14\% and 39\% Late Yields and only 4\% and 5\% Early Yields, respectively. 
This may be explained by the fact that FP is trained on read speech, and while TT2 is trained on conversational speech the content is largely monologic, where a single speaker talks at length about a topic with minimal input from the interlocutor.

According to the naturalness scores, Microsoft outperforms all systems averaging a MOS of 4.8, while the least natural voice was TT2 with a score of 3.9. Microsoft also provides the most intelligible speech with a WER of 2.3\%, Amazon and Google are slightly worse with 2.7\% WER, while FP and TT2 are less intelligible, averaging 5.6\% and 5.2\%. 

\textbf{Text Manipulation}: 
Because TTS models are commonly trained on text that includes punctuation, we provide two different input permutations of each agent turn that can induce different prosodic realizations at the end of the SD sentence. 
Whereas a period denotes the end of a sentence, commas are used to separates parts within them, and can potentially condition the TTS systems to generate stronger hold cues. 
Filled pauses (or \textit{fillers} for short), such as ``um'', are also well known to be a strong cue to turn hold \cite{clark100117, Ball1975}. To study if such manipulations would indeed have a turn-holding effect, we experiment by replacing the ending period with either a comma or with the filler ``um,'' (also ending with a comma). 

The comma prompt produced an increase of Weak Holds, for all models (except TT2), and raised the Weak Hold frequency to over 90\% for each system except Google (83\%), as seen in Table \ref{tab:metrics_comma}. 
The FP model showed the largest relative improvement, classifying 56\% of the pauses as a Strong Hold. Microsoft had the second largest gain, improving its Strong Hold classification from 20\% to 33\%, outperforming Google (29\%), but falls short of Amazon with 48\%.
The large improvement for FP could be due to it being trained on read speech, where the source text is grammatical and contains a lot of punctuation, making the prompt conditioning prominent. 
However, TT2 was largely unaffected by the change of prompt which, similarly, could be because it is trained on spontaneous speech, that naturally contains less punctuation, making it less sensitive to punctuation in general.
The results indicate that while the effect of changing periods to commas provides stronger hold signals, the strength varies depending on the systems. Furthermore, given the simplicity and the straight-forward meaning of punctuation control it could be beneficial for spontaneous TTS to include such examples in their training data.


\begin{table}[th]
  \caption{The aggregate metrics for the COMMA permutation. All values are percentages (\%) except for the MOS.}
  \label{tab:metrics_comma}
  \centering
  \begin{tabular}{lccccc}
    \toprule
    \textbf{Metric}&\textbf{AMZN}&\textbf{GGL}&\textbf{MSFT}&\textbf{FP} &\textbf{TT2}\\ 
    \midrule
    Weak Hold$\uparrow$& 92 & 83 & 90 & 90 & \textbf{97} \\
    Strong Hold$\uparrow$  & 48 & 29 & 33 & 54 & \textbf{92} \\
    Early Yield$\uparrow$& \textbf{39} & 28 & 28 & 6 & 5 \\
    Late Yield$\uparrow$& 92 & \textbf{94} & 89 & 43 & 18 \\
    \midrule
    MOS$\uparrow$& 4.2 & 4.3 & \textbf{4.8} & 4.4 & 3.8\\
    WER $\downarrow$& 2.8 & 2.8 & \textbf{2.2} & 5.4 & 6.1\\
    \bottomrule
  \end{tabular}
\end{table}

Interestingly, while the manipulation only changes the punctuation prior to the pause, the later yield signals are affected as well. 
The cause of this effect can be because of differences in the generated speech at the end of the questions or how the entire turn is perceived. 
Because the VAP model is trained on \SI{20}{\second} segments of continuous dialog, it can learn to utilize context over multiple sentences to infer the yield probabilities. 
However, the effect is small across all systems and does not change the general outcome in a significant way. 
Lastly, we note that both the intelligibility and naturalness metrics are roughly unaffected by the manipulation. 

While changing the punctuation does not introduce any additional words or letters to the prompt, inserting a filler does. Whereas a filler can be considered to function as a word \cite{clark100117}, it is not guaranteed that all TTS systems include them in their training, and the realization of their characteristic hesitation-like properties \cite{fox_tree_interpreting_2002} can vary between systems. Table \ref{tab:metrics_filler} shows that both Microsoft and TT2 achieve 100\% Strong Hold when a filler is included. While TT2 shows strong hold cues in both the original and comma permutations, the filler drastically strengthens the hold for the Microsoft voice. From listening to the generated filler samples we note that both Microsoft and TT2 have learned to generate fillers in a prosodically relevant way. This is reasonable considering the spontaneous nature of the TT2 training data and distinguishes Microsoft from the other commercial services. These results show that if a TTS has the ability to produce fillers, it can be a very efficient way to avoid user barge-ins. 

\begin{table}[th]
  \caption{The aggregate metrics for the FILLER permutation. All values are percentages (\%) except for the MOS.}
  \label{tab:metrics_filler}
  \centering
  \begin{tabular}{lccccc}
    \toprule
    \textbf{Metric}&\textbf{AMZN}&\textbf{GGL}&\textbf{MSFT}&\textbf{FP}&\textbf{TT2}\\ 
    \midrule
    Weak Hold$\uparrow$& 93 & 77 & \textbf{100} & 92& \textbf{100} \\
    Strong Hold$\uparrow$& 57 & 26 & \textbf{100} & 69& \textbf{100} \\
    Early Yield$\uparrow$& \textbf{38} & 27 & 23 & 6 & 5 \\
    Late Yield$\uparrow$& 92 & \textbf{95} & 88 & 38 & 21\\
    \midrule
    MOS$\uparrow$& 4.2 & 4.2 & \textbf{4.7} & 4.3 & 3.8\\
    WER $\downarrow$& 6.5 & 4.9 & \textbf{3.1} & 7.5 & 7.2 \\
    \bottomrule
  \end{tabular}
\end{table}

\textbf{Prosody Manipulation}: 
As an alternative to text manipulation, we 
experiment with direct manipulation of the prosody prior to the pause, to achieve prosodic signals that are associated with turn-hold, according to the literature \cite{Skantze2021,tt_cues}. 
We exclusively target the last syllable of the SD sentence, corresponding to the last word given the text extraction step, and raise the intensity, make the duration longer and flatten the intonation. 
We consider two approaches where we (1) apply post-processing on the original generated speech, using Praat \cite{Praat} and torchaudio, 
and (2) use the FP and TT2 innate ability to control for these properties explicitly in the generation step.

The post-processing produces a substantial increase of Strong Holds for both Microsoft (81\%) and Amazon (70\%), while Google (53\%) is less affected, see Table \ref{tab:metrics_fsh}.
This indicates that Google provides more ambiguous signals earlier on in the statement and requires greater intervention to produce consistent turn-holding aspects over the course of the turn.
While TT2 consistently produces strong hold cues, FP shows a relative improvement similar to Microsoft and increases its Strong Hold performance to 85\%. 
Overall, the results highlight the possibility for TTS systems to improve their generative capabilities w.r.t. the functional turn-taking aspects of conversational speech. 

\begin{table}[t]
  \caption{The aggregate post-processing metrics. All values are percentages (\%) except for the MOS.}
  \label{tab:metrics_fsh}
  \centering
  \begin{tabular}{lccccc}
    \toprule
    \textbf{Metric}&\textbf{AMZN}&\textbf{GGL}&\textbf{MSFT}&\textbf{FP}&\textbf{TT2}\\ 
    \midrule
    Weak Hold$\uparrow$& \textbf{97} & 92 & 96 & 97 & \textbf{99} \\
    Strong HOLD$\uparrow$& 70 & 53 & 81 & 85 & \textbf{98}\\
    Early Yield$\uparrow$& \textbf{46} & 26 & 29 & 6 & 4 \\
    Late Yield$\uparrow$& 95 & \textbf{97} & 95 & 42 & 18\\
    \midrule
    MOS$\uparrow$& 3.8 & 3.9 & \textbf{4.6} & 4.0 & 3.6\\
    WER $\downarrow$ & 2.7 & 2.5 & \textbf{2.4} & 3.9 & 5.2 \\
    \bottomrule
  \end{tabular}
\end{table}
Post-processing inadvertently introduces artifacts that affects the perceived naturalness, reflected by the consistent decrease of the MOS across all TTS systems.
The alternative approach, controlling the prosody directly in the generation process, alleviates this negative impact, where FP produces MOS of 4.3 instead of 4.0 and TT2, 3.7 instead 3.6, as compared to the post-processing. 
However, the hold signals are less prominent and the Strong Hold score for TT2 (93\%) is unaffected (as compared to the original), while FP produces the same score as for the simpler comma prompt (54\%).

\section{Conclusion}
We introduced a new automatic evaluation method that can measure a TTS model's ability to produce turn-taking cues. 
We show that although commercial TTS systems often do provide appropriate turn-taking cues, they still produce ambiguous, or opposite, cues for up to 21\% of pauses (Google) and up to 10\% of yields (Microsoft).
Furthermore, we show that TTS models trained on read or spontaneous speech are generally good at producing turn-holding cues but show low performance w.r.t. turn-yielding cues.
Replacing a period with a comma is a simple approach to condition TTS models to provide stronger turn-holding cues. 
If a TTS model has the ability to correctly generate fillers, they can be added to a prompt to strongly convey the agent's intention to continue its turn.
By directly manipulating the prosody, we show the possibility of improvement for conversational TTS to produce turn-taking cues independent of the provided input prompt.

While controllable TTS has mainly focused on prosodic control \cite{lancucki2021fastpitch}, or style control (such as emotion or speaker) \cite{wang2018style}, we argue that controlling for turn-taking cues could provide added benefits useful for conversational speech. 
A VAP model could even provide turn-taking signals to be optimized during training, that could enable the ability to generate appropriate turn-taking cues without relying on additional datasets.

This work is breaking down barriers by providing a free, open-source, model that can be used to evaluate the generative turn-taking capabilities of conversational TTS systems. 
It enables researchers with limited resources to conduct fast model iteration, without relying on expensive human evaluations, and is a step towards making speech research more equal in general.

\section{Acknowledgements}
This work was supported by the Riksbankens Jubileumsfond (RJ) project (P20-0484), the Swedish Research Council projects (2020-03812) and (VR-2019-05003), and finally the Digital Futures project, \textit{AAIS}.

\bibliographystyle{IEEEtran}
\bibliography{mybib}
\end{document}